\documentstyle[epsf]{mn}

\title[Clusters \& Constraints]
      {Constraining our Universe with X-ray \& Optical Cluster Data}

\author[J.M Diego et al.]
   { J.M  Diego$^{1,2}$, E. Mart\'\i nez-Gonz\'alez$^1$, J. L.
    Sanz$^1$, L. Cay\'on$^1$, J. Silk.$^3$ \\
   $^1$Instituto de F\'\i sica de Cantabria, Consejo Superior de
     Investigaciones Cient\'\i ficas-Universidad de Cantabria,
     Santander, Spain\\
   $^2$Departamento de F\'\i sica Moderna, Universidad de Cantabria,
     Avda. Los Castros s/n, 39005 Santander, Spain\\ 
   $^3$ Department of Physics. Nuclear \& Astrophysics Laboratory,
        Keble Road, Oxford OX1 3RH, UK\\}

\pagerange{\pageref{firstpage}--\pageref{lastpage}}

\begin{document}

\maketitle

\label{firstpage}

\begin{abstract}
We have used recent X-ray and optical data in order to impose some 
constraints on the cosmology and cluster scaling relations.\\
Generically two kind of hypotheses define our model. 
First we consider that the cluster 
population is well described by the standard Press-Schechter (PS) formalism, 
and second, these clusters are assumed to follow scaling 
relations with mass: Temperature-Mass ($T-M$) and X-ray Luminosity-Mass 
($L_x - M$).\\
In contrast with many other authors we do not assume specific 
scaling relations to model 
cluster properties such as the usual $T-M$ virial relation or an 
observational determination of the  $L_x-T$ relation. 
Instead we consider general unconstrained parameter scaling relations. \\
With the previous model (PS plus scalings) we fit our free  
parameters to several X-ray and optical data sets with the advantage 
over preceding works, that we consider all the data sets at the same time. 
This prevents us from being inconsistent with some of the available observations. 
Among other interesting conclusions, we find that only 
low-density universes are compatible with all the data considered and that 
the degeneracy between $\Omega_m$ and $\sigma_8$ is broken. 
Also we obtain interesting limits on the parameters characterizing the scaling 
relations.\\

\end{abstract}

\section{Introduction} \label{intro}
In recent years, the quality and quantity of new data sets coming from 
several X-ray missions allow a more precise study of the properties of 
galaxy clusters. These data, together with optical data sets have 
allowed many 
authors
to compare the predictions of different models with 
observations. \\
The standard approach is to simulate the data for a given parameter 
dependent model and then by using an estimator 
(likelihood, $\chi ^2$, etc) look for the 
best fitting model. That is, the best parameter combination which best 
fits the data. \\
In
this process usually several assumptions are made. 
The most usual is that concerning the cluster population. Normally it is 
assumed 
that the cluster population is well described by the 
Press-Schechter (PS) formalism (Press \& Schechter 1974). 
This approach is supported by 
N-body numerical simulations which do show a good agreement with the 
PS parameterization 
(Efstathiou et al. 1988; White et al. 1993; Lacey \& Cole 1994; 
Borgani et al. 1999).\\

\noindent
Another assumption usually made is the scaling of the temperature 
of the cluster with its mass, the $T-M$ relation, which is taken 
as the virial relation (Eke et al. 1996). 
A $T-M$ relation is necessary, for instance to 
build the temperature function of clusters (see section \ref{model}).
However, it is not clear to what extent the virial assumption is true for 
clusters, especially for those at high redshift. Several works show that 
the relation between mass and temperature has an exponent close or equal 
to the virial exponent $M \propto T^{\frac{3}{2}}$ 
(Evrard et al. 1996; Horner et al. 1999; Neumann \& Arnaud 1999). 
However, the isothermal $\beta$-model and 
X-ray surface brightness deprojection masses follow a steeper 
$M \propto T ^{1.8-2.0}$ scaling (Horner et al. 1999).\\
There are other scaling relations which are not well understood in the sense 
that they depend on the data used to build those relations and also on the 
method used to fit the data. A good example of this point is the 
Luminosity-Temperature relation ($L_x-T$). From the literature one can find 
scaling relations ranging from $L_x \propto T^{2.6}$ (Markevitch 1998) to 
$L_x \propto T^{3.3}$ (David et al. 1993) while the most common one is 
$L_x \propto T^{2.9}$  
(White et al. 1997; Arnaud \& Evrard 1999; Reichart et al. 1999). 
They show a discrepancy in the exponent of the relation.
More and better data will 
be needed
to resolve that discrepancy. 
Fabian et al. (1994) noted that this scatter is mostly due to clusters 
with strong cooling flows. See also White et al. (1997) for a good 
discussion about the effect of cooling flows. 
Also the method used to fit the $L_x-T$ data 
can explain part of this scatter. Conventional least-squares regression 
analysis assumes the abscissae data have zero error. This problem is 
overcome, for instance, by the use of an algorithm that takes into account 
errors in both dimensions of the data (White et al. 1997).\\

\noindent
At present,  
X-ray observations are 
the best available data to study clusters.
The amount of available X-ray data is increasing 
fast and in the
near future larger data sets will be available.
The strong X-ray emission from the hot gas in the
intracluster medium makes the X-ray surveys an ideal way to detect clusters
of galaxies. New catalogues of clusters have been published in the last
years with the advantage that they are X-ray selected, and new ones
are in preparation.\\

\noindent
Clusters have been used to impose constraints on cosmology in several 
papers (Oukbir \& Blanchard 1992; Lupino \& Gioia 1995; 
Eke et al. 1996; Donahue 1996; 
Kitayama \& Suto 1997; Oukbir \& Blanchard 1997; 
Mathiesen \& Evrard 1998; Donahue \& Voit 1999; and many others).
Clusters are the largest gravitationally bound objects in the universe
and represent the final stage of the peaks in the primordial matter
distribution. Their distribution in the mass-redshift (M,z) space is
the fingerprint of those primordial fluctuations.
The cluster abundance and its evolution is an essential cosmological 
test. Their modelling only depends on cosmological parameters and not on 
any cluster scaling relation like the $T-M$ or $L_x-T$, thereby allowing 
a more precise determination of the cosmological parameters independently 
of any assumption about the cluster scaling relations.
For this reason many authors have tried to determine
the cluster mass distribution as a function of redshift, the mass function
(Bahcall \& Cen 1993; Biviano et al. 1993; Girardi et al. 1998).
These authors found many difficulties when they tried a direct determination
of the mass function. Basically, the problem is that the mass estimators 
are usually based on different assumptions (spherical symmetry, 
virialization, hydrostatic equilibrium). Lensing determinations
work pretty well but the number of clusters with mass determined by
this technique is too small to build a mass function from them.\\
An improvement could be to compare the models with the data using 
other X-ray derived functions (luminosity, flux, temperature). 
The advantage 
of using X-ray data is that the determination of 
the luminosity, flux or temperature of the clusters is in general 
less affected by sy errors than the usual mass determination 
based on radial velocities of galaxies. \\

\noindent
In this paper we want to extract some information about 
clusters and 
cosmological parameters from cluster data. 
Our aim is to find a model (PS plus scalings) which fits 
different observational data. 
This model will be realistic in the sense that it describes present 
observations (mass, temperature, and X-ray luminosity and flux functions).\\
This work follows many others but with two 
main differences. First, in our model we will allow a large number of 
free parameters (9) instead of the one or two free parameter models 
usually assumed. 
This will prevent us from doing wrong assumptions about the scalings 
$T-M$ or $L_x-M$ which could affect the final conclusion. 
Our second difference is that we will consider 
different data sets simultaneously. 
This is an important point as we will show in section \ref{data}, 
where we demonstrate how 
some models with a good fit to some data sets, are however inconsistent with 
others.  \\
The structure of the paper is the following. In section \ref{data}, 
we describe the different data sets which will be used in the fits, 
and in section \ref{model} we describe the model used to fit the 
previous data. In section \ref{best_fit}, we search for the 
best model fitting the different data sets and discuss the best model 
estimator. In section \ref{discussion}, we discuss the main results and 
compare them with previous works. Finally, section \ref{conclusions} 
includes the main conclusions of this paper and some implications for 
future  X-ray \& CMB experiments.\\

\noindent
Trough this paper we assume $H_0 =  100 h$ km s$^{-1}$ Mpc$^{-1}$. 
Although we work in $h$ units, the previous assumption should 
be taken into account when comparing with other results.
\section{The data} \label{data}
In this work we have compared our model (Press-Schechter and 
$T-M$ and $L_x-M$) with five different data sets. 
$dN(M)/dM$ (Bahcall \& Cen 1993), 
$dN(M,z)/dM$ (Bahcall \& Fan 1998), $dN(L_x)/dL_x$ 
(Ebeling et al. 1997) , 
$dN(S_x)/dS_x$ (Rosati et al. 1998; de Grandi et al. 1999), 
and $dN(T)/dT$ (Henry \& Arnaud 1991).\\

\noindent
The first one is the mass function given in Bahcall \& Cen (1993) which is 
built from a compilation of optical data of nearby clusters ($z < 0.1$). 
These data have several uncertainties mainly 
due to the poor precision in the determination of cluster masses. They 
estimated the masses through the richness and velocity dispersion of the 
clusters. More sophisticated methods, as lensing estimation would be 
preferable in order to achieve a good mass function but unfortunately 
the number of clusters with masses estimated from gravitational lensing 
is too small. 
It is important to  bear in mind that masses in Bahcall \& Cen (1993) 
where obtained from proportionality laws between cluster richness and mass 
or velocity dispersions and mass. 
Therefore, these masses estimates should be considered as inferred masses 
and not as a {\it direct} measure. 
There are other more recent determinations of the mass 
function (Girardi et al. 1998) but they suffer from the same problems. 
From the theoretical point of view, the mass function has the advantage 
of depending only on the cosmological parameters and not 
on the parameters in the $T-M$ or $L_x-M$ relations. 
Therefore the mass function is very useful to constrain the 
cosmological parameters.\\
We would like to point out that the original mass function given in Bahcall 
\& Cen (1993) is a cumulative mass function. We have computed the 
differential mass function from the previous one by computing the 
difference between consecutive bins and the corresponding error bars are 
build from the original ones by adding them quadratically.
Also important is to note that in Bahcall \& Cen (1993) masses are estimated 
within a radius of $1.5 h^{-1} Mpc$. Our masses however are estimated 
within the virial radius. 
As a first approximation we will consider that the mass 
within a sphere of $1.5 h^{-1} Mpc$ centered on the cluster and 
the virial mass are equivalent. This is justified because  
virial radius can be well approximated by 
$r_v = 1.3 M^{1/3}(1+z) h^{-1} Mpc$ which, for typical clusters, 
is of the order of $1$ $h^{-1} Mpc$. Clusters with masses  
$M < 1.5 \times 10^{15} h^{-1} M_{\odot}$ will have virial radius below 
$1.5 h^{-1} Mpc$. 
In our model we have considered a truncated cluster density profile 
beyond the virial radius. Therefore, the previous clusters will have the 
same masses for larger radii ($1.5 h^{-1} Mpc$).
Some problems could arise with very massive clusters 
with $M > 1.5 \times 10^{15} h^{-1} M_{\odot}$ 
but these ones are very rare and the correction factor  will be in any case 
small.\\

\begin{figure}
   \begin{center}
   \epsfxsize = 9.cm
   \begin{minipage}{\epsfxsize}\epsffile{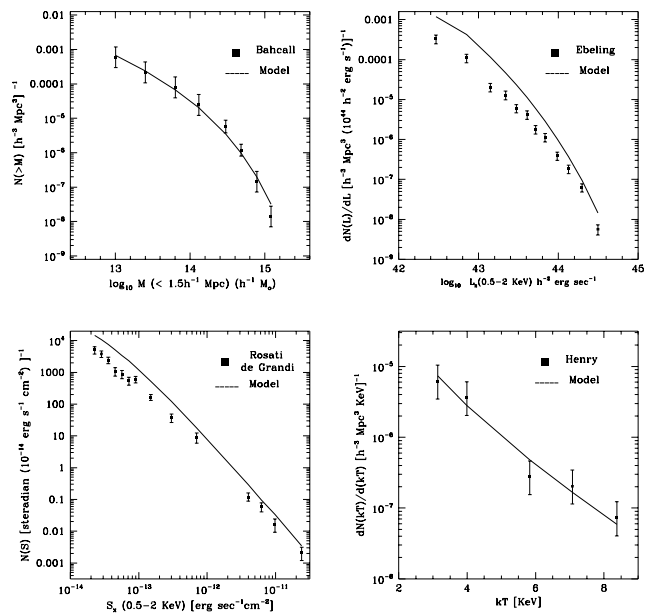}\end{minipage}
   \caption{ \label{fig_bad} An example of a bad model. 
            The fit is good in the case 
            of the mass and temperature functions but this model does not 
            reproduce the other two curves. This model has the following 
	    typical values 
            for the parameters which are commonly used in the 
	    literature (see text in subsection \ref{Scaling} and 
	    section \ref{discussion} for an 
	    explanation of these parameters and a discussion of their values): 
            $\sigma_8 = 0.8 , \Gamma = 0.2, 
            \Omega_m = 0.3, (\Lambda = 0), T_0 = 1.0\times 10^8  h^{\alpha} K, 
            \alpha = 2/3, \psi = 1.0, 
            L_0 = 1.0\times 10^{45} h^{\beta} h^{-2} erg/s, 
            \gamma = 2.9, \phi = 3.0$ } 
   \end{center}
\end{figure}

\noindent
In order to account for the evolutionary effects in the mass function, we have 
also considered another data set: the evolution of the mass function 
for massive clusters (Bahcall \& Fan 1998). In this data set, the error 
bars are large but the data are good enough to constrain the cosmology
even more. 
Bahcall \& Fan have demonstrated that combining the two data 
sets can impose strong constraints on $\Omega_m$ and $\sigma_8$.\\
Obviously the best models found by Bahcall \& Fan should be compatible with 
other data sets but we have shown that this point is not true in general. 
If we take models with a good fit in both, the mass function and the 
evolution of the mass function, we have found that only a few of 
those models have also a good fit in other different data sets 
(for example the luminosity function, see fig. \ref{fig_bad}). 
This is the main reason why we decided to work with several 
presently available cluster data sets at the same time. 
We looked for the model that simultaneously fits the different 
data sets the best. The additional data sets came from X-ray observations.\\

\noindent
Several X-ray cluster catalogues have been published recently 
(Rosati et al. 1995; Burke et al. 1997; Collins et al. 1997; 
Scharf et al. 1997; Ebeling et al. 1998; Vikhlinin et al. 1998; 
de Grandi et al. 1999; Voges et al. 1999; Romer et al. 2000 
and references therein). 
Some of these catalogues are deeper in flux than others and 
they have different sky coverages. The techniques to detect the clusters are 
also different (wavelets, Vikhlinin et al. 1998;  
Voronoi tesselation and percolation, Ebeling \& Wiedenmann 1993), 
but they show a remarkable agreement in the results. 
Particularly remarkable is the good agreement in the
luminosity function among all those works, showing that the 
estimation of the luminosity function is a robust indicator of the 
cluster population and this function will be very useful in the 
process of fitting our model.
For the luminosity function we have used the estimation of Ebeling et al. 
1997. This luminosity function is built from a ROSAT $90 \%$ flux-complete 
sample of $\sim 200$ bright clusters (Brightest Cluster Sample, BCS) 
in the northern hemisphere at 
high galactic latitudes ($ \vert \delta \vert \geq 20^{\circ}$), 
with measured redshifts $z \leq 0.3$ and 
fluxes higher than $4.4 \times 10^{-12} erg cm^{-2} s^{-1}$ 
in the 0.1-2.4 KeV band.
Different determinations of the luminosity function have been given in 
the literature 
(Burke et al. 1997; Rosati et al. 1998; Vikhlinin et al. 1998), 
being all of them compatible with that in Ebeling et al. (1997). We would like 
to point out that this curve is given for an Einstein-de-Sitter Universe 
with $q_0 = 0.5$. In order to build the luminosity function is 
necessary to assume 
a cosmological model for the computation of the luminosity distance and 
the comoving volume. We have checked the effect of changing $\Omega_m$ in this 
function. 
We have seen that the effect is negligible when we are 
dealing with redshifts below 0.3 as in this case. For higher redshift data, 
the effect is still small as it can be appreciated from fig. (\ref{fig_bad}) 
of Bahcall \& Fan (1998) where the authors show the data for three different models. \\
Furthermore there are other functions that 
can be used as a test of our model. In particular, the flux function  
is relatively well established (there is only a small scatter among 
the different author estimations). 
The main difference between these two functions is the 
redshift and cosmological model assumed. The flux function is a 
{\it direct} measure in the sense that this function does not contain any 
information about the distance (redshift plus cosmological model) from which 
the cluster is emitting. 
On the contrary, the luminosity function contains this additional 
information (redshift plus cosmological model). 
Both functions are obviously connected by the assumed model.

\noindent
For the flux function we used the one given by Rosati et al. (1998) for 
low-flux clusters and for the bright part of the curve we used the function 
of De Grandi et al. (1999). The sample of Rosati et al. (1998) 
(ROSAT Deep Cluster Survey, RDCS) is over the 
redshift range $0.05-0.8$ and is a complete flux-limited subsample of 70 
galaxy clusters, representing the brightest half of the full sample, 
which have been spectroscopically identified down to the flux limit 
$4 \times 10^{-14}  erg cm^{-2} s^{-1}$ in the 0.5-2.0 KeV band. 
In the RDCS sample, the sky coverage is small (48 deg$^2$) 
meanwhile the sample of de Grandi et al. has a larger sky coverage 
(8235 deg$^2$) but the limiting flux is higher
 ($\sim 3.5 \times 10^{-12} erg cm^{-2} s^{-1}$ in the 0.5-2.0 KeV band) 
and therefore the sample is shallower ($z \leq 0.3$) than the RDCS sample.\\

\noindent
The final curve we have used to constraint our model is the temperature 
function. Henry \& Arnaud (1991) compiled a temperature function from a 
sample of 25 nearby clusters. Their sample is  
X-ray selected and comes from Lahav et al. (1989) subject to the 
additional restrictions that the flux in the 2-10 KeV must be 
$\geq 3 \times 10^{-11}  erg cm^{-2} s^{-1}$ and the galactic latitude 
($ \vert b^{II} \vert \geq 20^{\circ}$) (see Piccinotti et al. 1982).  
The sample is greater than $90 \%$ complete and redshifts range between 
$z = 0.0036$ and $z = 0.09$.\\
The temperature function of Henry \& Arnaud (1991) is known to suffer 
from some errors (Eke et al. 1996, Markevitch 1998, Henry 2000) but as 
mentioned in Eke et al (1998), and Henry (2000) the errors in the 
Henry \& Arnaud (1991) temperature function are largely compensated. 
The temperature function is usually presented in integral form. A 
determination of the differential temperature function requires binning 
the data and performing an average over the objects in the bin. This 
procedure introduces some arbitrariness that the integral form avoids. 
However, due to the fact that our method is based on $\chi^2$ quantities we 
need the temperature function in a differential form. The arbitrariness 
of this binned function could be reduced significantly by increasing the 
number of clusters with measured temperature. However, there are few clusters 
for which we know precisely their temperature and consequently the differential 
temperature function is poorly determined. 
In order to check the validity of the Henry \& Arnaud (1991) temperature 
function with more recent data we computed a binned version of the 
temperature function using the Henry (2000) data. 
Our estimate of the differential temperature function showed to be 
in good agreement, within the error bars, with the previous estimate 
of Henry \& Arnaud (1991). 
Due to this agreement and to the large error bars of this function, 
our results will not depend significantly on the choice of one or 
another temperature function.\\
Although the temperature function is affected by large error bars, 
however its use is justified because as a difference with the 
luminosity or flux functions, only the $T-M$ relation is needed 
to build the temperature function. To compute the theoretical luminosity and 
flux functions from the PS formalism, the $L_x-M$ and $T-M$ relations 
are needed. The first one is used to obtain the bolometric luminosities 
from the mass and the second is required to obtain the luminosities 
in the observed band. 
Hence, the temperature function is less 
affected by the uncertainties in the cluster scaling relations than 
the luminosity and flux functions. A recent determination of the 
temperature function can be found in Blanchard et al. (2000) and Henry (2000).
Their determination of this function is compatible with 
the one in  Henry \& Arnaud (1991) for temperatures $> 3$ KeV. 

\noindent
The information about the redshift and sky coverages, limiting flux, and 
the energy band in which luminosities and fluxes are given is needed in order 
to correctly simulate the data following the characteristics of 
the observations.
The total number of clusters, and thereby, the error bars, will depend 
on the redshift and sky coverages and also on the limiting fluxes. 
The shape of the functions will depend  
on the limiting flux because lowering the limiting fluxes  
less massive and more distant clusters will be selected.  
Energy band and $K$ corrections must also be included in order 
to correct for the bolometric luminosity. Finally 
the cluster number densities are based on the computation of the $V_{max}$
which is the maximum volume in which the cluster could have been and 
still remained in the sample. Therefore these volumes will depend on the 
the limiting flux 
(see Page \& Carrera 2000 for a good discussion about the $1/V$ method).\\
All those observational features will be considered to perform a bias test 
using Monte Carlo simulations of the models in section \ref{best_fit}.  \\

\noindent 
These data sets are not completely independent. Some clusters are 
common to the different catalogs and one should consider the dependence 
between the data but it can be shown that the dependence is not very 
significant. 
The luminosities and fluxes are independent because to compute the 
luminosity from the flux the redshift is needed.
 Because the redshift 
is an independent variable with respect to the flux, then the luminosity 
should be also considered as independent with respect to the flux. 
The temperature is another independent quantity so we do not expect 
correlations between this data set and the others.
However,
there is a clear correlation between the first data point in the evolution 
of the mass function and one point of the local mass function. 
Indeed the information given by the comoving number density of clusters 
$N(M > 8.0\times 10^{14} h^{-1} M_{\odot})$ at $z=0$ 
is contained in both data sets. 
Apart from this, we consider that the rest of our data points 
are in fact independent. \\
The situation is different with the theoretical curves. The model will 
introduce some correlations among the curves, as we will see in the 
next section.

\section{The model} \label{model}
\subsection{The Press-Schechter formalism}
As 
in previous 
works the starting point of our model is the mass function 
which contains the information about how many clusters are at a given redshift 
and how massive they are.
We adopt the standard Press-Schechter formalism (Press \& Schechter 1974) 
which has shown to be very consistent with N-body simulations 
(Lacey \& Cole 1994; Borgani et al. 1999). \\
In this formalism the cluster number density per unit mass 
as a function of mass and redshift is given by:
\begin{equation}
 \frac{dN(M,z)}{dV(z)dM} = \sqrt{\frac{2}{\pi}}\frac{\bar{\rho}}{M^2}
 \frac{\delta _{co}(z)}{\sigma _M} \left \vert\frac{dlog\sigma _M}{dlogM} 
 \right\vert  e^{-\frac{\delta _{co}(z)^2}{2\sigma _m^2}} , 
\end{equation}
\noindent
where $\bar{\rho}$ is the present day average matter density 
$\bar\rho = \Omega_m \times 2.7755 \times 10^{11}  h^2 M_{\odot} Mpc^{-3}$ 
and $\delta _{co}(z)$ is the linear theory overdensity extrapolated at 
the present time for a uniform spherical fluctuation collapsed at 
redshift $z$.\\
For a $\Omega_m = 1$ model we have used 
$\delta _{co}(z) = 1.6865(1 + z)$ and for 
$\Omega_m < 1$ we take $\delta _{co}(z) = \frac{D(0)}{D(z)}\delta _c(z)$ 
where $D(z)$ is the linear growth factor (Peebles 1980)  and :
\begin{equation}
\delta _c(z) = \frac{3}{2}D(z)
\left( 1 + \left(\frac{2\pi}{sinh(\eta) - \eta}\right)^{2/3} \right ) ,
\end{equation}
\noindent
for an open $\Lambda=0$ model and :
\begin{equation}
\delta _c(z)=1.6866[1+0.01256log_{ 10}\Omega_m(z)] ,
\end{equation}
\noindent
for a flat $\Lambda$CDM model 
(see Kitayama \& Suto 1996, Mathiesen \& Evrard 1998 for details). \\ 
$\sigma_M$ is the rms of the density fluctuation at the mass scale M which is 
related with the power spectrum of density fluctuations $P(k)$ through :
\begin{equation}
  \sigma_M^2 = \frac{1}{2\pi}\int _0^\infty dk k^2 P(k) W^2(kR) ,
\end{equation}
\noindent
where the window function $W(kR)$ is introduced in order to select the volume 
from which the object with mass M will be formed.
We have used the standard top hat approach for the window 
function and the corresponding Fourier transform is in this case:
$W(kR) = 3 (sin(kR) - (kR) cos(kR))/(kR)^3$. R is the comoving scale 
corresponding to the mass M and the relation between both quantities is :
$M = \bar\rho \frac{4}{3}\pi R^3$.\\
For the power spectrum we have used the following parameterization,
\begin{equation}
 P(k) = A\sigma_8^2 k^n T^2(k) .
\end{equation}
The amplitude $A$ is computed from equation (4) just taking in that 
equation the mass corresponding to $R = 8$ $h^{-1}$ Mpc and 
eliminating from both sides of the equality the parameter $\sigma_8$. 
$n$ is the primordial power spectrum. We fixed this parameter to the 
Harrison-Zeldovich case $n=1$ according to determinations 
from CMB data (COBE-DMR Bennet et al. 1996; MAXIMA Balbi et al. 2000), 
and finally $T(k)$ is the transfer function. 
For the transfer function we used the fit given in Bardeen et al. 
(1986) for an adiabatic CDM model:
\begin{eqnarray}
T(k)&=&\frac{ln(1+2.34q)}{2.34q} \times  \\
    & & [1+3.89q+(16.1q)^2+(5.46q)^3+(6.71q)^4]^{-1/4} \nonumber ,
\end{eqnarray} 
\noindent
where $q=k(h Mpc^{-1})/\Gamma$, being $\Gamma$ the shape parameter 
of the power spectrum. For the case of a CDM model with negligible 
$\Omega_b$, then $\Gamma \sim \Omega_m h$. 
We have considered as an additional constraint  
in our calculations the following. Although all our data sets 
and quantities are $h$ independent (everything is in $h$ units), 
however we have just considered 
those models for which the ratio $\Gamma / \Omega_m$ is between the 
conservative limits  $0.5 < h < 0.75$, thus avoiding 
to compute 
CDM models which could be inconsistent with recent determinations of $h$.\\

\noindent
In the previous formalism, there are two main variables: the mass and redshift 
of the cluster. Therefore, the Press-Schechter mass function which predicts  
the density of clusters expected at a given redshift and mass  
can be considered as the probability distribution of clusters
in the mass-redshift space (M-z) by normalizing by the total number. 
The cosmological parameters in this formalism are basically 
three, the density of the universe, $\Omega_m$, the amplitude of the power 
spectrum which we parameterize in units of $\sigma_8$ and finally the shape 
parameter of the power spectrum $\Gamma$.\\
We can compare this model with real observations of the mass function 
and by doing this we can get 
some information about these three parameters. This has been done in several 
works (Bahcall \& Cen 1993; Girardi et al. 1998) and the conclusions are very 
interesting. These works have shown for instance that low-density 
universes are more compatible with the observed mass function.\\
However, there are some problems with these works. First, the quality 
of the data is not very good, mainly due to the fact that 
most of the masses have been estimated using radial velocities 
of cluster galaxies.
Second, the mass functions are built only for nearby clusters and 
these mass functions do
not contain any information about the cluster abundance 
at high redshift. There are some attempts to estimate the evolution of 
the mass function with redshift and, although the error bars are 
very large, one can obtain very interesting constraints on the 
cosmological parameters using this evolution (Bahcall \& Fan 1998).
This indicates that an accurate information of the cluster abundance at 
high redshift would be a very powerful technique to constrain the 
cosmology. Unfortunately the mass function of clusters at high redshift 
is not well determined yet but there are some other functions which can be 
used in addition to the mass function. Recent X-ray experiments (Einstein, 
ASCA, ROSAT) have determined the temperature, luminosity and flux for 
several hundreds of clusters, some of them at medium and high redshift 
(up to $z=0.8$ in the RDCS). 
This information can be used to build new functions 
similar to the mass function, based on the temperature, luminosity 
and flux of the clusters. For instance 
the 
expected temperature function up to a given redshift, $dN/dT$,  
(which can be compared 
with the corresponding observational temperature function), 
will be given by the integral along the redshift interval of: 
\begin{equation}
 \frac{dN(T,z)}{dV(z)dT} =\frac{dN(M,z)}{dV(z)dM}\frac{dM}{dT},  
\label{correl}
\end{equation}
where $dN(M,z)/dV(z)dM$ is the Press-Schechter mass function. 
In order to build that function we need to calculate the 
derivative  $\frac{dM}{dT}$ and hence a $T-M$ relation is required. 
Usually the virial relation is assumed; $T \propto M^{2/3}(1 + z)$, 
though as discussed below, we will introduce free parameters to
describe this relation . 
To build the X-ray luminosity and flux functions we operate in the same way 
but in this case we need the relation between the mass and the X-ray 
luminosity of the cluster, the $L_x-M$ relation. 
There are few attempts to determine observationally the $L_x-M$ 
relation but the situation is different with the $L_x-T$ relation 
(David et al. 1993; Markevitch 1999; Reichart et al. 1999). 
These works show that there is a scaling in this relation 
$L_x \propto T^{2.6 - 3.3}$. 
The exponent of the scaling depends on whether or not clusters with cooling 
flows are considered, being the exponent higher when clusters with 
cooling flows 
enter  
the analysis. 
Another contribution to that scattering is that different statistical methods 
have been used to analyze the data (White et al. 1997). 
 
\noindent
Using the $T-M$ relation and the $L_x-T$ scaling is possible to 
build an $L_x-M$ relation which can be used to construct the 
luminosity and flux functions.\\
\subsection{Cluster scaling relations}\label{Scaling}
Starting from the Press-Schechter mass function plus the 
$T-M$ and $L_x-M$ relations, the idea of this work is, 
therefore, to build the mass function itself 
and the remaining curves: temperature, X-ray luminosity and flux 
functions. 
We will 
compare 
these curves
with 
the 
corresponding observational data sets and by changing our model 
parameters we 
will look 
for the best model simultaneously compatible 
with all the different data sets.\\
So, all what we need to know are the $T-M$ and $L_x-M$ relations. \\
For the $T-M$ relation, the most common model comes from the virial theorem 
plus the spherical collapse model and the isothermal gas distribution 
assumption (Eke et al. 1996):
\begin{equation}
 T_{gas} \propto M_{vir}^{\frac{2}{3}}(1 + z) .
\end{equation}
The shortcomings of this relation are well known 
(Eke et al. 1996;  Kitayama \& Suto 1996; Viana \& Liddle 1996; 
Voit \& Donahue 1998). Basically 
the problem is that this assumption only holds for virialized objects. 
In the case of clusters this is more or less true for low redshift clusters 
where the equilibrium conditions required by the virial theorem are achieved. 
But we do not know what happens at high redshift. 
Similar problems are in the redshift evolution of this relation. As discussed 
in Voit \& Donahue (1998), 
the consequences of using an inaccurate $T-M$ relation 
can be quite significant. For these reasons, we 
will consider this relation as an unconstrained one  
and we will adopt as the $T-M$ relation the following, with no previous 
assumption about the parameters:
\begin{equation}
  T_{gas} = T_0 M_{15}^{\alpha}(1 + z)^{\psi} ,
  \label{Tx}
\end{equation}
where $M_{15}$ is the cluster mass in $h^{-1} 10^{15} M_{\odot}$. 
For the $L_x-M$ relation the situation is similar. The $L_x-M$ relation is not
well established and we prefer to allow this relation to be a free parameter 
relation,  
\begin{equation}
  L_x^{Bol} = L_0 M_{15}^{\beta}(1 + z)^{\phi} .
  \label{Lx}
\end{equation}
\noindent
Since $T_{gas}$ is in Kelvin and $L_x^{Bol}$ in $h^{-2} erg/(s\  cm^2)$ 
and considering the mass in $h^{-1} 10^{15} M_{\odot}$, 
then an additional $h^{\alpha}$ and $h^{\beta}$ 
must be introduced in $T_0$ and $L_0$ respectively in order to make our 
result $h$-independent. \\
From the previous $L_x-M$ relation it is possible to build the 
$S_x-M$ relation by simply considering, 
\begin{equation}
  S_x^{Bol} = \frac{L_x^{Bol}}{4 \pi D_l(z)^2 } .
  \label{Sx}
\end{equation}
\noindent
In this formalism the $L_x-T$ relation has the form: 
\begin{equation}
  L_x^{Bol} = \frac{L_0}{T_0^{\gamma}} T^{\gamma}(1 + z)^{\delta} \  , 
\label{L_T} 
\end{equation}
\noindent
where $\gamma = \beta / \alpha$ is the familiar exponent of the $L_x-T$ 
relation and $\delta = \phi - \psi \beta/\alpha$. 
Within this framework we have a total of 9 free parameters: 
 $\sigma_8, \Gamma, \Omega_m, T_0, \alpha, \psi, L_0,\beta$ and $\phi$ 
(or equivalently we can use $\gamma = \beta/\alpha$ instead of $\beta$, and  
$\delta = \phi - \psi \gamma$ instead of $\phi$). 
We have also considered the two 
situations flat $\Lambda = 1 - \Omega_m$ models ($\Lambda$CDM) 
and open $\Lambda=0$ models (OCDM). \\
There are some experimental determinations of the parameters in $T-M$ and 
$L_x-M$. For instance many works have shown that $\alpha$ 
is compatible with the predicted virial 
value $\alpha = 2/3$ (Neumann \& Arnaud 1999) but also possible are scaling 
exponents $\alpha \sim 0.5$ (Horner et al. 1999; Nevalainen et al. 2000). 
The normalization of the $T-M$ scaling has been determined by many 
authors and they found typical values of $T_0 \sim 1.0\times 10^8 K$
(Horner et al. 1999). There is not too much work done 
on the determination of the redshift exponent $\psi$ because the 
data and redshift coverage is poor to fit this exponent, but usually what 
is found is that this exponent is also compatible with the virial 
prediction $\psi \approx 1$ (Neumann \& Arnaud 1999). 
On the $L_x-M$ relation the scatter in the 
data is too large (large error bars in mass) but the situation gets better 
when the $L_x-T$ relation is instead considered. In the latter case, 
the scatter in the correlation is reduced. 
Typical values for the parameters in these relations are 
$L_0 \sim 1.0 \times 10^{45} h^{-2} erg/s$ , 
$\gamma \sim 2.9$ (Arnaud \& Evrard 1999) and $\delta \sim 0$ 
(Borgani et al. 1999; Reichart et al. 1999; Fairley et al. 2000) 
although the uncertainty in this last parameter is large. From 
the relation between $\delta$ and $\phi$ is easy to infer that 
$\phi \approx 3$ is what it is expected when $\psi = 1$ and $\gamma \sim 3$.
In fig. (1) the model was chosen according to these typical values. 
From $L_x-T$ and $T-M$ is easy to infer the parameters in $L_x-M$ and 
vice-versa. \\
In our fit, we have allowed the parameters to take different values around 
these observational and theoretical predictions.\\

\noindent
We are now ready to build the theoretical 
five curves $dN(M)/dM$, $dN(M,z)/dM$, $dN(L_x)/dL_x$, $dN(S_x)/dS_x$, 
and $dN(T)/dT$ and to look for the
best model by comparing these curves with 
the data.\\
Similar analysis have been presented in previous works.
However, we would like to remark 
again that in those works either some parameters are fixed (in $T-M$ or 
$L_x-M$) or only one data set is used (e.g. $dN(M)/dM$, $dN(T)/dT$, etc). 
In Mathiesen \& 
Evrard (1998), the authors combined a free parameter $L_x-M$ relation 
and two data sets ($dN(L)/dL$, and $dN(S)/dS$) in order to say something 
about the evolution of the $L_x-T$ relation. However, they fixed the $T-M$ 
relation and they 
did not combine together the results coming from the two different data sets. 
A similar work was done in Borgani et al. (1999) where the authors 
have used the observables, flux number counts, redshift distribution 
and X-ray luminosity function over a large redshift baseline ($z < 0.8$) 
of the RDCS in order to constrain cosmological models. 
In the same paper, no assumption is made {\it a priori} 
on the $L_x-M$ relation, 
except for the amplitude of this relation which is fixed by the authors. 
In addition the $T-M$ relation is fixed to the usual spherical 
collapse plus virial plus isothermal gas distribution model.\\
In Bridle et al. (1999) they have combined the X-ray 
cluster temperature function 
(Henry \& Arnaud 1991, Henry 2000) with CMB data and the IRAS 1.2 Jy galaxy 
redshift survey, but they have assumed a fixed $T-M$ relation.
This latter point can affect the final result.\\
Up to now, no previous work has combined such a large number of data sets 
as the five ones we have used without including
any assumptions about the 
normalization or specific scalings of the temperature or X-ray luminosity. \\

\noindent
As we mentioned at the end of the previous section, the model we have assumed 
will introduce some correlations between the 5 theoretical curves. Just by 
looking to equation (\ref{correl}), 
it is clear that the temperature function is correlated with the mass function 
(equivalently for the luminosity and flux functions). This point should be 
taken into account when fitting the data.
\section{Statistical analysis and results} \label{best_fit}
In order to fit the five data sets we must decide which estimator 
we should use.
Because we assume there are some scaling relations between mass and 
temperature ($T-M$) and luminosity ($L_x-M$) in the X-ray band, 
then, there must be some correlations among the five simulated data sets. 
Therefore we should start by considering an estimator like the standard 
likelihood estimator which takes into account all the correlations into 
the correlation matrix $M$.
In our case, the model depends on 9 free parameters and if we consider a 
grid of, let's say 5 values per parameter, then we should compute the 
correlation matrix for $5^9 \sim $ 1 million different models. This process 
would take many years. A faster technique would require a search method 
that 
avoids to explore all the parameter space. 
This could be the technique if we were interested just in the best model 
but we also want to know the error bars, or in other words the 
probability distribution of the parameters. 
To do that we need to know the probability in a given regular grid.\\
To simplify the problem, the most simple approach 
is to consider 
the standard $\chi^2_{joint}$
as our estimator:
\begin{equation}
 \chi_{joint}^2 = \chi_M^2 + \chi_{M(z)}^2 + \chi_{T}^2 +  
                  \chi_{L_x}^2 + \chi_{S_x}^2 ,  
\end{equation}
\noindent
where $\chi_i^2$ represents the corresponding ordinary $\chi^2$ for the 
five different data sets and we are assuming that the correlation matrix 
is in this case diagonal.\\
By doing this, we know that we are forgetting the correlations between 
the curves and that there will be some bias in our estimation. For this 
reason, we want to check other more elaborated estimators.\\
We have considered as a second estimator of the best model 
one based on Bayesian theory (Lahav et al. 1999);
\begin{equation}
 -2 ln P_L = \chi_{L}^2 ,
 \label{Lahav}  
\end{equation}
\noindent
where, 
\begin{equation}
 \chi_{L}^2 = \sum_i^5 N_i ln(\chi_i^2) . 
 \label{Lahav2}
\end{equation}

\begin{figure}
   \begin{center}
   \epsfxsize=9.cm
   \begin{minipage}{\epsfxsize}\epsffile{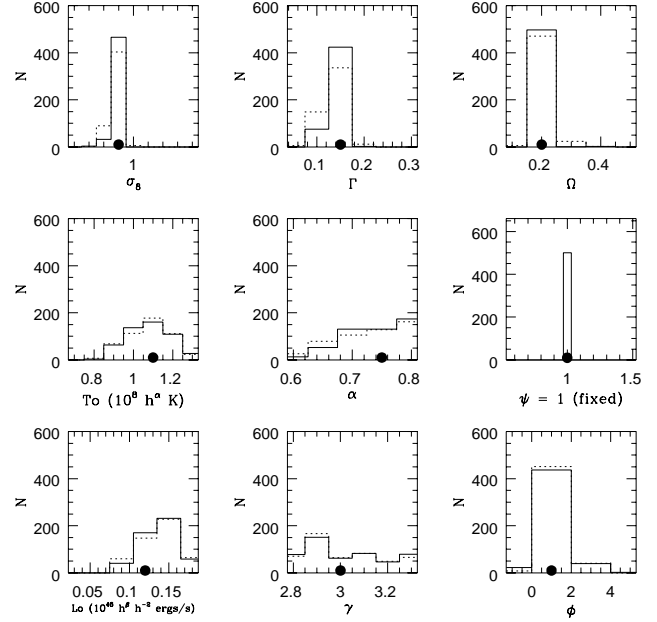}\end{minipage}
   \caption{\label{fig_bias} 
            The histogram represents the number of times each parameter 
            was considered as part of the best model by the standard 
            $\chi^2_{joint}$ (dotted) and the $\chi^2_{L}$ (solid). 
            The black dot represents the input model.}
   \end{center}
\end{figure}

\begin{table}
\large
\caption{Best $\Lambda$CDM  and OCDM models ($\psi$ fixed to 1.0).
Error bars represent the projection of the 
contour at the 68 \% confidence level of the 8-dim probability on each 
of the parameters.
Limits marked with ($*$) must be considered as lower limits because 
the parameter was not explored above that limit.}
\label{table}
\normalsize
\begin{tabular}{lcccc}
\hline
\hline
Parameter  & OCDM & $\Lambda$CDM \\
\hline
\hline
$\sigma_8$     		      & $0.8^{+0.1}_{-0.1}$ & $0.8^{+0.2}_{-0.1}$\\
\hline
$\Gamma$      		      & $0.2^{+0.05}_{-0.1}$ & $0.2^{+0.05}_{-0.1}$\\
\hline
$\Omega_m$     		      & $0.3^{+0.2}_{-0.1}$ & $0.3^{+0.2}_{-0.1}$\\
\hline
$T_0 (10^8 h^{\alpha} K)$     & $1.1^{+0.1}_{-0.2}$ & $1.1^{+0.2}_{-0.3}$\\
\hline
$\alpha$       		      & $0.8^{+0*}_{-0.15}$ & $0.75^{+0.05*}_{-0.1}$\\
\hline
$\psi$        		      & $1.0             $ & $1.0$\\
\hline
$L_0 (10^{45}h^{\beta}h^{-2}erg/s)$  & $0.9^{+0.6}_{-0.0}$ & $1.5^{+0.3}_{-0.9}$\\
\hline
$\gamma$  		      & $3.1^{+0.2*}_{-0.3}$ & $3.2^{+0.1*}_{-0.4}$\\
\hline
$\phi$			      & $3.0^{+0}_{-2}$ & $1.0^{+2}_{-0}$\\
\hline
\end{tabular}
\label{table_best}
\end{table}
\noindent

\noindent
In this estimator, the $\chi_i^2$ is again the ordinary $\chi^2$ 
for each data set and $N_i$ represents the number of data points for 
the data set $i$. Based on a Bayesian approach with the choice 
of non-informative uniform priors on the log,  
those authors have seen that this estimator is appropriate for the case when 
different data sets are combined together, as is our case. The factor $N_i$ 
plays the role of a weight factor. Larger data sets are considered 
more reliable for the parameter determination.\\
We have checked both estimators by performing
a bias test. In this test we have 
simulated the five data sets for a concrete model with the corresponding 
error bars 
similarly as they were computed in the real data. 
The input model was selected according to the criterion that it would be as 
close as possible to the data (for instance the model which minimizes 
$\chi^2_{joint}$). 
In the simulations, we have taken into account all the characteristics of 
the data, that is, sky coverage, limiting flux, maximum redshift, etc.
Then we compare each one of these realizations corresponding to the assumed 
model with the models previously computed in the grid and
for each realization we get the best-fitting model to the simulated 
data using both estimators. \\
In fig. \ref{fig_bias}, we plot the number of times each parameter 
was considered as part of the best model by the first and second estimator. 
The dot represents the input model. 
As it can be seen from the histograms the second estimator $\chi_L^2$ works  
a bit better than the standard $\chi_{joint}^2$. There is still some bias but 
the agreement between the input model and the recovered peak of the 
distribution is very good. \\
We can get some interesting information from these plots. 
The dispersion 
of the histograms indicates how sensitive is the estimator to that parameter.
For instance, the cosmological parameters are well constrained. This is 
not the case for the redshift exponent $\psi$. 
We fixed this parameter to the virial value $\psi=1$ because our method is 
not sensitive to that exponent. When changing this exponent the simulated 
curves did not change appreciably, showing the almost null dependence of the 
simulated curves to this parameter. 
There is an explanation to that. This exponent appears 
only in the $T-M$ relation as the redshift exponent. This relation is needed 
to construct the temperature function and these data goes only up to 
redshift $\sim 0.1$. Therefore, it is not surprising 
that we can not get any significant result about the redshift dependence 
with these data. 
The $T-M$ relation appears also in the calculation of the 
X-ray luminosity in a given band, so the exponent $\psi$ would be 
in principle important when we are simulating clusters at high z to compare the flux 
function with the data of RDCS since this data goes to $z \sim 0.8$. 
The flux in the band used by Rosati et al. (1998) is calculated 
from the luminosity in that band (see eq. \ref{Sx}) and $L_{band}$ is 
computed from $L_{Bol}$ in the following way : 
$L_{band} = L_{Bol} f_{band}$ 
where 
$f_{band}$ includes the band and $K$ corrections and is 
usually well approximated by the integral of 
the frequency dependence of the Bremsstrahlung emission : 
$f_{band} = - \frac{1}{K_bT} \int_{E_{min}}^{E_{max}} e^{ -\frac{E}{K_bT} }dE$.
$E_{min}$ and $E_{max}$ 
are
the energy limits of the band, and $T$ the cluster 
temperature. The redshift dependence of $f_{band}$ is concentrated in the 
K-correction, and there is a weak dependence also on the redshift exponent 
of the $T-M$ relation.
This dependence is too weak to be able to impose 
some constraints on this exponent even when we are using data at medium-high 
redshift like the fluxes of clusters at $z \sim 0.8$ (Rosati et al. 1998).
This 
explains 
the reason
why with these data we can not say much about 
the exponent $\psi$. 
We decided to fix this parameter to the 
standard value $\psi = 1$, therefore reducing our dimension in the 
parameter space from 9 parameters to 8. However, this parameter should 
be considered as a free parameter when dealing with future data for which the 
redshift coverage will increase significantly.\\
Other 
result
from the bias test is that there is some bias in the 
parameter $\alpha$ (scaling exponent of the T-M relation). The bias is 
about 0.05 or more towards higher values of $\alpha$. We will come later to 
this point. A similar bias is found in $L_0$ (about 0.03 to higher values). 
The bias is not too large considering the small bin interval but anyway 
it must be taken into account.\\
Apart from these parameters, the second estimator seems to be a good 
indicator of the best model. \\

\begin{figure}
   \begin{center}
   \epsfxsize=9.cm 
   \begin{minipage}{\epsfxsize}
         \epsffile{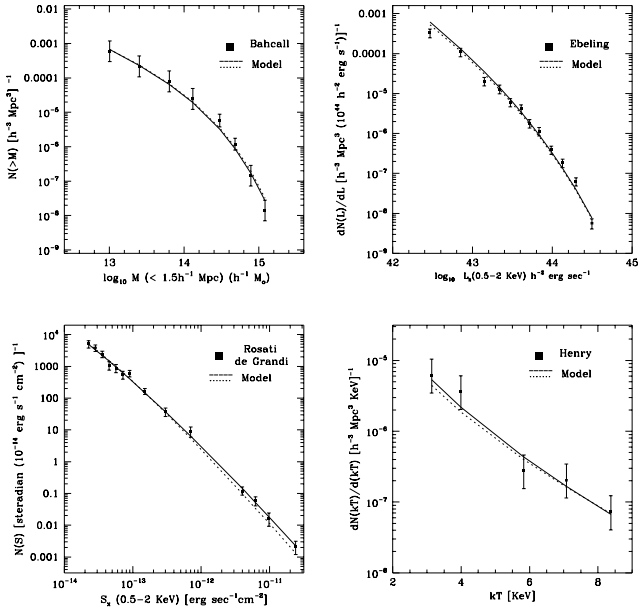}\end{minipage}
   \caption{\label{fig_best_4curves}
            Expected curves compared with data for the best $\Lambda$CDM 
            model (solid) and  OCDM model (dotted). See 
            table \ref{table_best}}
   \end{center}
\end{figure}

\begin{figure}
   \begin{center}
   \epsfxsize=8.cm 
   \begin{minipage}{\epsfxsize}
         \epsffile{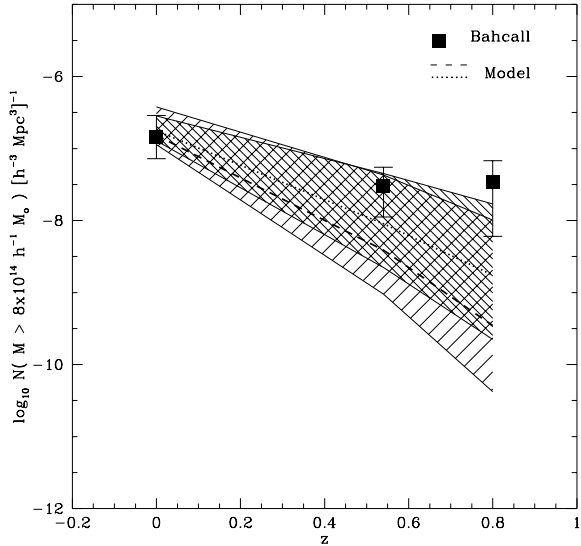}\end{minipage}
   \caption{\label{fig_best_Mz}
            The best $\Lambda$CDM (dashed) and OCDM (dotted) 
            models compared with the fifth data set (evolution of 
            the mass function). Although the model is out the 
	    $68$\% error bars, however, although not represented, 
	    the model is inside the $95$\% error bars (see Bahcall \& Fan 
	    (1998)). The shaded regions correspond to the $68$\% confidence 
            region of the cosmological parameters (high dense shaded OCDM, 
            and low dense shaded $\Lambda$CDM)}
   \end{center}
\end{figure}

\noindent
The next step 
is 
to compute the probability distribution in our 9 dimension parameter space 
(8 after fixing $\psi$), using the second estimator. 
We have used a grid with about 2 million different models in the two cases 
flat $\Lambda$CDM and OCDM and for each of them we have 
computed its $P_{L}$ (eq. \ref{Lahav}). 
\\
In fig. \ref{fig_best_4curves}, we plotted the best model compared 
with four data curves used in the fit.
It is important at this point to compare figure 1 and figure 3. 
Both cases only differ slightly on the cluster scaling relations 
but the differences in the models are relevant, specially in the case 
of the luminosity and flux functions. This shows the sensitivity  
of the models to  the cluster scaling relations. Small 
changes in the parameters of these scalings can produce a 
completely different function if all the changes imply variations for the 
function in the same direction. 
The best models listed in table 1 are an example 
of a {\it fine tunning} between the parameters. One change in one parameter 
should be compensated by another change in other(s) parameter(s) in order to 
keep the model compatible with the data and only a small region of the  
parameter space is allowed. This also explains why the temperature 
function does not change significantly. 
While in the luminosity and flux functions 
both scaling relations ($T-M$, and $L_x-M$) are needed, in the case of the 
temperature function only the $T-M$ relation is required, thereby reducing 
the number of parameters and consequently the change in the 
temperature function when a variation in the whole set of parameters is 
performed.\\
In fig. \ref{fig_best_Mz}, the best model is compared with the fifth curve. 
There is a  good agreement between our 
best-fitting model and all the data sets except the fifth one 
where the model predicts less comoving number densities at high $z$ 
than observed (only 2 clusters in the $z \sim 0.54$ bin and 1 in the 
$z \sim 0.8$ bin). However, one should bear in mind that 
in the fifth curve there are only three data points and also these data 
points have large error bars and therefore the weight of the fifth curve in 
the Lahav et al.'s estimator (see eq. \ref{Lahav2}) is low compared with 
the weight of the other data sets. 
When considering the band corresponding to the $68$\% confidence 
region of the cosmological parameters, it overlaps the data within the $68$\% 
error bars.\\
On the other hand, the $dN(M,z)/dM$ curve is useful in the 
sense that including this curve in the analysis, helps to break  
the degeneracy between $\sigma_8$ and $\Omega$ 
(as we will show in the next section).\\
Obviously, this point suggests the need of getting better quality data in the 
evolution of the mass function in order to make these data a decisive 
discriminator between the models. \\

\section{Discussion}\label{discussion}
\begin{figure}
   \begin{center}
   \epsfxsize=9.cm
   \begin{minipage}{\epsfxsize}\epsffile{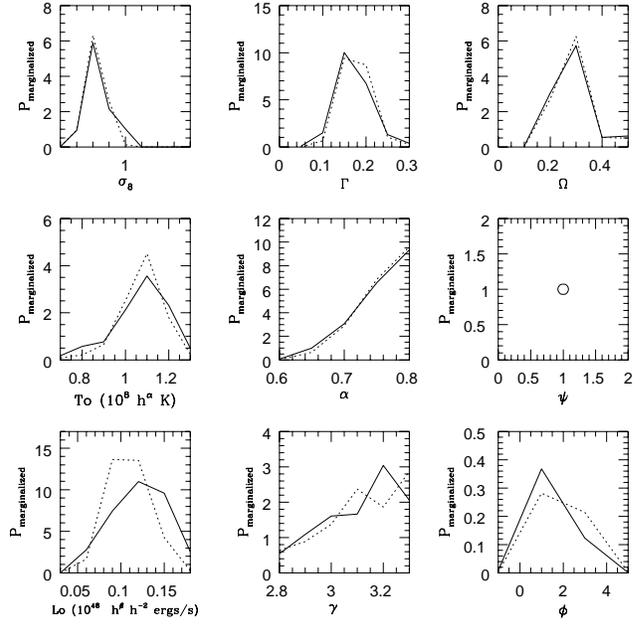}
   \end{minipage}
   \caption{\label{fig_margin}
            Marginalized probability distributions for the 8 parameters. 
            Dotted lines for open CDM models and solid 
            lines for flat $\Lambda$CDM models. In both cases the $\psi$ 
            parameter was fixed to 1 (see text).}
   \end{center}
\end{figure}

We have computed the marginalized probability of the parameters 
in order to see how well constrained are those parameters. In fig. 
\ref{fig_margin}, we show the power of the method to constrain the 
cosmology, even the amplitudes of the $T-M$ and $L_x-M$ relations 
are well constrained. 
As 
seen
in the bias test, it is clear 
that we 
can not say much about the exponents $\alpha$ and $\gamma$, except that 
high values are favored.\\
Virial theory predicts $\alpha = 2/3$ which is compatible (at 68 \%) with 
our fit values given in table \ref{table_best}. 
However, models with $\alpha = 0.8$ work better 
than virial models, and maybe higher values could work even better.
(We did not check this possibility because we wanted to remain within values 
of the parameters not far away from the expected ones). 
In Nevalainen et al. (2000), the authors found 
$\alpha \sim 0.55$ which is inconsistent with the self-similar (virial) 
prediction. They argue that a possible explanation for this discrepancy 
is preheating of intracluster gas by supernova-driven galactic winds 
before the clusters collapse, as proposed by e.g. David et al. (1991), 
Evrard \& Henry (1991), Kaiser (1991) and Loewenstein \& Mushotzky (1996).
If supernovae release a similar amount of energy per unit gas mass 
in hot and cool clusters, the coolest clusters would be affected more 
significantly than the hottest ones.
This increase in their temperature 
will change the slope in the $T-M$ relation towards low $\alpha$ 
values. In the data sets we have considered, we have bright clusters 
with temperatures which are typically $T > 3$ KeV. At those 
high temperatures, the previous effect should not be relevant and hence 
the slope in the $T-M$ relation should approximate the self-similar 
value ($2/3$) (see fig. 2 in 
Nevalainen et al. 2000). This can explain how our results are more 
compatible with the virial prediction that with those empirical relations 
where cool clusters are included in the fit.\\
A possible source of systematic errors in our best fitting values 
(including $\alpha$) can be on the data themselves. 
The data sets used in this work suffer from several systematics 
which can affect the best fitting parameters in the $T-M$ relation. 
In our method, the best fitting $T-M$ relation is obtained from a 
global fit of the model to all the data. 
If such data sets change in some way then the best fitting model should 
change as well. In the mass function, masses are 
defined inside a fixed radius. A different choice of this fixed radius could 
produce a different estimate of the cluster mass function. In the X ray 
flux and luminosity functions, the inferred fluxes and luminosities depend 
on the assumed cluster profile used to extrapolate the observed surface 
brightness profile (Vikhlinin et al. 1998).
If masses, fluxes or luminosities are underestimated or overestimated, 
then we should expect some differences in the best fitting parameters and 
in particular in $\alpha$.\\
These systematics will be reduced with future determinations of these 
quantities ($M,T,L_x$). Cluster mass estimates can be clearly improved using 
the lensing technique. On the other hand, on-going X ray missions (CHANDRA, 
Newton-XMM) will be able to determine the cluster surface brightness profile 
at larger radii and with a higher quality for a significant number of 
clusters.
Furthermore, from the bias test we know that in the $\alpha$ parameter 
there is some bias in the peak of the distribution, so we know that if we 
got $\alpha = 0.8$, this high value compared to the virial one can be due 
to the bias in our estimator. 
However, our estimate of $\alpha$ is compatible (given the error bars) with 
the virial exponent. 
It could also be that hot clusters really behave 
in this way, showing a tendency towards high $\alpha$ exponents.
In order to distinguish between the two possibilities, 
more and better quality data is needed.\\
The second exponent, $\gamma$, is also {\it pointing} to high values. 
In this case we know, from the bias test, that this exponent is 
degenerated. This together with the error bars found can very well 
accommodate an exponent $\gamma \sim 2.9$, 
which is the most frequent value obtained in the literature when fitting 
directly the $L_x-T$ relation. 
However, the direct estimate of $L_x-T$ suffers from large 
scattering and depending 
on the kind and number of clusters considered the results are quite 
different and high values for $\gamma$ should not be ruled out yet.\\
For instance, in Borgani et al. (1999) they found $3 < \gamma < 4$ when 
fitting a phenomenological $L_x-T$ relation plus PS to the local X-ray 
luminosity function.\\  
Concerning the redshift exponent $\phi$, we have a bit more 
information compared with the null information we got in $\psi$. 
This is not surprising 
because the $L_x-M$ relation appears in the calculation of $dN(L)/dL$ and 
$dN(S)/dS$ where the data is between $z \in [0,0.3]$ and $z \in [0,0.8]$ 
respectively, and these redshift intervals are much deeper than the one 
for the $dN(T)/dT$ data. Although the best value differs for the two 
cosmologies considered, however the value $\phi \sim 3$ is allowed in 
both cases. Experimentally, there is no determination of the $\phi$ 
parameter. What the different authors assume
when they try to fit 
the $L_x-M$ relation to real data, is that there is no redshift 
dependence in this relation, that is, they simply fit the relation 
$L_x = L_0 M^{\beta}$. 
However, we have shown in section \ref{data}, 
that the unobserved $\phi$ parameter can be related to the redshift 
exponent in the $L_x-T$ relation (eq. \ref{L_T}), 
$\delta = \phi - \psi\gamma$ and using this relation we can infer the 
value of $\phi$. Typical values for $\delta$ found in the 
literature are $\delta \sim 0$ (Fairley et al. 2000). 
In Borgani et al. (1999), the authors have shown that the $L_x-T$ relation
is compatible with no evolution. This result is also 
consistent with that of Mushotzky \& Scharf (1997) where they compared 
results from a sample of ASCA temperatures at $z > 0.14$ with the low 
redshift sample by David et al. (1993) and they found that data out to 
$z \simeq 0.4$ are consistent with no evolution in $L_x-T$. \\
Now if we   
assume
$\psi = 1$ (from virial models) and $\gamma \sim 3$ (from 
the empirical $L_x-T$ relation) then $\phi$ should be $\phi \sim 3$ 
in order to satisfy $\delta \sim 0$. So we can conclude that $\phi \sim 3$ 
is compatible with the virial assumption and also with $\gamma \sim 3$.\\

For a comparison of our results with a recent determination of 
the $L_x-T$ relation see for instance Fairley et al. (2000). 
It is remarkable that in that paper the 
authors find $\gamma = 3.15$, very close to our preferred value. Also 
they found an amplitude in the $L_x-T$ relation which is 
$C = 6.04 \pm 1.47 \times 10^{42}$ erg/s. 
This value should be compared with the 
amplitude $L_0$ in our $L_x-M$ relation $L_0 \approx 1.0 \times 10^{45} 
 h^{\beta} h^{-2}$ erg/s which corresponds to an amplitude 
in $L_x-T$ (see eq. \ref{L_T}) $L_0/T_0^{\gamma} = 6.25 \times 10^{42}$ erg/s 
(for $\gamma = 3$, $T_0 = 1.0\times 10^8 h^{\alpha}$ K and taking $h = 0.5$ 
which is the value used in Fairley et al. 2000).
The normalization obtained here for the $T-M$ relation is higher 
than those ones obtained from simulations or pure cluster modelling 
(spherical symmetry, virialization, hydrostatic equilibrium). 
This is not surprising as these kind of modelling does not include 
some physical processes relevant to cluster formation and evolution. 
Our results should be compared 
with observational determinations of this relation like the ones in 
Horner et al. (1999) where they found values for the $T-M$ normalization 
compatibles with our estimate (see table 1 in Horner et al. 1999).\\ 
\noindent
It is important to point out that not all the parameter combinations 
inside the error bars in table \ref{table_best} correspond to 
models which are simultaneously compatible with all the data sets. As 
we have shown in fig. 1, the model with parameters
$ \sigma_8 = 0.8 , \Gamma = 0.2,$ 
$\Omega_m = 0.3, (\Lambda = 0), T_0 = 1.0\times 10^8$ K, 
$\alpha = 2/3, \psi = 1.0, L_0 = 1.0\times 10^{45}  h^{\beta} h^{-2}$ erg/s, 
$\gamma = 2.9, \phi = 3.0$ is an example of a `{\it bad} ' model in the sense 
that this model does not fit all the data sets.
One should also notice 
that although these values are inside the error bars given 
in the table, since they are projected ones, not all the possible 
combinations are allowed at the $68$\% confidence level.
Therefore, when choosing a model it is important to bear in mind 
the correlations among the parameters.

\begin{figure}
   \begin{center}
   \epsfxsize=10.cm 
   \begin{minipage}{\epsfxsize}
         \epsffile{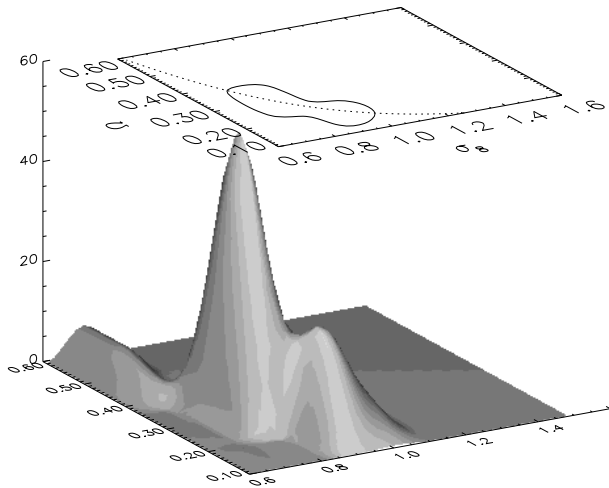}\end{minipage}
   \caption{\label{fig_Sigma8_Omega}
            Marginalized probability in $\sigma_8-\Omega_m$ for the flat  
            $\Lambda$CDM case (for OCDM the situation is similar). 
	    The probability distribution has been interpolated 
	    in order to smooth the surface. The contour shows the region at 
	    the $65$\% confidence level and the dotted curve corresponds to 
	    the correlation law: $\sigma_8 = 0.5\Omega_m^{-0.4}$.}
   \end{center}
\end{figure}

\noindent
The method is really powerful in the determination of 
the cosmological parameters. We made a consistent fit to five different 
data sets and we got strong constraints on the cosmological parameters. 
Independently of $\Lambda$, only low-density universes are compatible with 
the different data sets. The amplitude of the power spectrum is also 
well constrained. 
Its value is consistent with, for instance, the 
value obtained by Bridle et al. (1999) where they have combined cluster, 
plus CMB and IRAS data using the same Lahav et al.'s estimator and they 
obtained $\sigma_8 \sim 0.75$ and $\Omega_m \sim 0.35$.\\
We have computed the marginalized probability in the ($\sigma_8-\Omega_m$) 
space in order to look for the well known $\sigma_8 - \Omega_m$ correlation 
(Eke et al. 1996; Carlberg et al. 1997; Henry 1997; Kitayama \& Suto 1997; 
Bahcall \& Fan 1998; Borgani et al. 1999; Bridle et al. 1999). 
From the five data sets, the function $dN(M,z)/dM$ shows a tendency 
to favor low-density models ($\Omega \leq 0.2$) 
whereas the others seem to favor slightly higher values of $\Omega$.
Although our grid is poor (intervals of $0.1$ in $\sigma_8$ and 
$\Omega_m$), we have seen that by combining the five data sets, 
there is a clear peak at the position cell 
($\sigma_8 = 0.8$, $\Omega_m = 0.3$) in both $\Lambda$CDM and OCDM models. 
Approximately $50$\% of the marginalized probability volume is enclosed 
in that $0.1\times 0.1$ cell (see fig. \ref{fig_Sigma8_Omega}).\\
This is showing that the degeneracy between these two parameters can be 
broken by combining different data sets. \\
    From the 5 data sets considered in this work, the evolution of the 
    cluster population with redshift (Bahcall \& Fan, 1998) 
    is, in principle,  the most sensitive to the change in the cosmological 
    parameters.
    However that data set suffers from large error bars due to the small 
    number of clusters present at the high redshift bins.
    We made an additional test to check the weight of this data set
    in our fit. 
    We have {\it recomputed} the marginalized probability in 
    $\Omega - \sigma_8$, excluding from the fit the Bahcall \& Fan (1998) 
    data set. The result is very simular to the one shown fig. 
    \ref{fig_Sigma8_Omega}.  
    This demonstrates that with only the low redshift data sets it is 
    possible to break the degeneracies present when each one of the 
    individual data sets is analyzed separately.\\

\noindent
The fit to the flat $\Lambda$CDM model was a bit better than 
the one to the open model in the sense that 
the best-fitting $\Lambda$CDM model had a 
smaller $\chi^2_L$ (76.1 compared with 76.8). 
In order to compare both cases in a more realistic way we 
performed the following 
statistical test. Using 500 simulations of the OCDM model, for each of them 
we got the best model given by the $\chi^2_L$ estimator applied to both 
situations ($\Lambda$CDM  and OCDM models). The result was that 197 of the 
initial 500 OCDM simulations had a smaller $\chi^2_L$ in the flat model case 
and in the remaining 303 simulations the open case was preferred. 
This demonstrates that both cases are equally probable with this method.\\

\noindent
Obviously, the constraints given here will improve when 
new and high quality data will be available (CHANDRA \& XMM-Newton). 
The method proposed should be very useful when constraining the cosmology 
with the upcoming new data. \\

\section{Conclusions} \label{conclusions}
In this work, we have shown that our method, which combines 
different data sets for the cluster population,
is a powerful tool to constrain both, the cosmology 
and cluster scaling relations. \\
Our method is robust in the sense that neither assumptions about the cosmology 
nor specific cluster scaling relations are made a priori.\\
Despite the correlations in the theoretical curves, we have shown 
that with simple estimators (like the standard $\chi^2_{joint}$ and 
the Lahav et al.'s Bayesian estimator) it is possible to fit the data 
without any significant bias. 

The main conclusions of this paper are the following.
Regarding the cosmology we have shown that only low-density (flat and open) 
models are compatible with the data sets considered in this paper. 
The marginalized probability in the ($\sigma_8-\Omega_m$) space shows a clear 
peak at the position ($\sigma_8 = 0.8$, $\Omega_m = 0.3$) in both 
$\Lambda$CDM and OCDM models. This is a very interesting conclusion 
because previous works 
(Eke et al. 1996; Kitayama \& Suto 1997; Bahcall \& Fan 1998; 
Borgani et al. 1999; Bridle et al. 1999) show a 
degeneracy in these two parameters. This degeneracy is broken when considering 
the five data sets we used in this paper. It is important to remark 
that in Bridle et al. (1999) the authors combine cluster abundance, 
CMB and IRAS data and they find values for ($\sigma_8,\Omega_m$)
very close to our best-fitting model. It is important to note 
that this result is compatible with the 
recent determination of the $\Omega_m$ parameter obtained by the BOOMERANG 
team (De Bernardis et al. 2000; Lange et al. 2000) and MAXIMA 
(Hanany et al. 2000; Balbi et al. 2000).\\

\noindent
The third cosmological parameter, $\Gamma$, is consistent with the value 
obtained from the fit of the power spectrum of galaxies assuming CDM. 
(Peacock \& Dodds 1994, Viana \ Liddle 1996)\\

\begin{figure}
   \begin{center}
   \epsfxsize=9.cm 
   \begin{minipage}{\epsfxsize}\epsffile{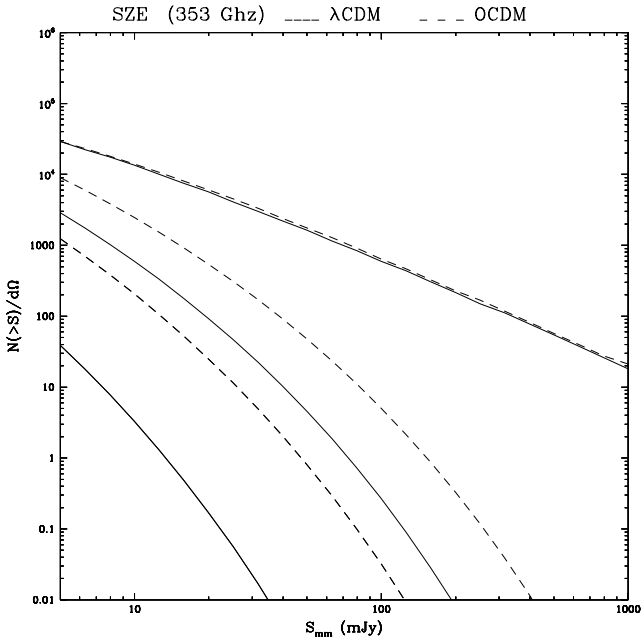}\end{minipage}
   \caption{\label{fig_SZE}
            Prediction of the integrated number counts of cluster 
            population al mm wavelengths (353 GHz) (SZE) for the best 
            flat (solid) and open (dotted) models in table 
            \ref{table_best}. In the plot three redshift shells are 
            represented: top $z < 1$, middle $z \in [1,2]$ and 
            bottom $z > 2$.}
   \end{center}
\end{figure}

\noindent
Regarding the parameters obtained for the cluster scaling relations, 
they are consistent with empirical determinations of such scalings. 
However, we find a tendency to high values in the $\alpha$ exponent 
which could contradict recent determinations of such exponent, 
Nevalainen et al. (2000). 
However, as mentioned in the discussion, we know that there is a bias in our 
estimation of $\alpha$. Therefore our estimate is compatible 
(within the error bars and the bias) with the virial exponent $\alpha = 2/3$.\\
Additional data coming from high redshift clusters (CHANDRA, 
XMM-Newton, PLANCK) will improve this result.\\
Particularly interesting is the work that can be done with future CMB surveys.
The PLANCK satellite will explore the whole sky at different 
frequencies (from 30 Ghz to 800 Ghz) and with resolutions between 5 arcmin 
and 30 arcmin. At these frequencies and with those resolutions we have shown 
(Diego et al. 2000) that many clusters are expected to 
be observed at high redshift ($z > 2$) through the Sunyaev-Zel'dovich effect 
(see fig. \ref{fig_SZE}). PLANCK is expected to detect those clusters 
with $S_{mm} > 30$ mJy. 
The information these clusters will provide will be decisive 
to definitely exclude many models.  
As shown for instance in Barbosa et al. (1996),  Aghanim et al. (1997),  
Diego et al. (2000), 
the SZE can be considered as a clear probe of the  
cosmological parameters. In particular, from the previous discussion we 
concluded that we are not able to discriminate between 
$\Lambda$CDM and OCDM models. 
However, from fig. \ref{fig_SZE}, it is evident that through the 
SZE it could be possible to distinguish between these two models 
at a very high confidence level.

\section{ Acknowledgments}
We would like to thank to Piero Rosati for kindly providing his data for the 
differential flux function and Nabila Aghanim for useful comments. 
This work has been supported by the 
Comisio\'on Conjunta Hispano-Norteamericana de Cooperaci\'on 
Cient\'\i fica y Tecnol\'ogica ref. 98138, Spanish DGESIC Project 
no. PB98-0531-C02-01, FEDER Project, no. 1FD97-1769-C04-01 and the 
EEC project INTAS-OPEN-97-1992. \\
JMD acknowledges support from a Spanish MEC fellowship FP96-20194004.
And finally JMD, EMG, JLS, and LK thank to the CfPA and Berkeley 
Astronomy Dept. for its hospitality during several stays in 1999.  

\newpage



\end{document}